\documentclass[aps,prd,preprintnumbers,superscriptaddress,nofootinbib,floatfix,twocolumn,notitlepage]{revtex4-2}

\usepackage[utf8]{inputenc} 
\usepackage[T1]{fontenc} 
\usepackage{lmodern}
\usepackage[british]{babel}
\usepackage{graphicx}
\graphicspath{{figs/}} 
\usepackage[pdfusetitle]{hyperref}

\usepackage{microtype} 
\usepackage{xfrac} 
\makeatletter
\g@addto@macro\@floatboxreset\centering
\makeatother

\usepackage[intlimits]{amsmath} 
\usepackage{amssymb} 
\usepackage{bm} 
\usepackage{bbm} 
\numberwithin{equation}{section} 
\allowdisplaybreaks 

\usepackage[capitalise]{cleveref} 
\creflabelformat{equation}{(\textup{#2#1#3})} 

\usepackage{booktabs} 
\usepackage{tabularx} 
\usepackage{longtable} 
\usepackage{array} 

\usepackage[italic]{hepnames} 

\usepackage{braket} 
\usepackage{slashed} 
\usepackage{siunitx} 
\DeclareSIUnit\invps{\ps^{-1}}
\usepackage{suffix} 

\newcommand{\Lagrangian}{\ensuremath{\mathcal{L}}\xspace}
\newcommand{\hc}{\text{h.c.}}

\newcommand{\V}[1]{\ensuremath{V_{#1}^{}}}
\WithSuffix\newcommand\V*[1]{\ensuremath{V_{#1}^*}}

\newcommand{\OHqone}{O^{(1)}_{Hq}}
\newcommand{\CHqone}{C^{(1)}}

\newcommand{\OHqthree}{O^{(3)}_{Hq}}
\newcommand{\CHqthree}{C^{(3)}}
\newcommand{\CHqthreebracket}{C^{(3)}}
\newcommand{\OHud}{O_{Hud}}

\newcommand{\program}[1]{\texttt{#1}}
\newcommand{\flavio}{\program{flavio}\xspace}
\newcommand{\smelli}{\program{smelli}\xspace}
\newcommand{\wilson}{\program{wilson}\xspace}
\newcommand{\mmeft}{\program{MatchMakerEFT}\xspace}

\newcommand{\mub}{\mu_b}

\newcommand{\muNP}{\mu_\mathrm{NP}}
\newcommand{\bsll}{\ensuremath{b\to s\ell^+\ell^-}}

\usepackage{xcolor}

\usepackage{cancel}

\begin{document}
\preprint{IPPP/26/76}

\title{Explaining $b\to s\ell^+\ell^-$ with CKM Unitarity Violation? A No-Go Theorem}

\author{Andreas Crivellin}
\email{andreas.crivellin@icrea.cat}
\affiliation{Grup de F\'isica Te\`orica, Departament de F\'isica, Universitat Aut\`onoma de Barcelona, 
E-08193 Bellaterra (Barcelona), Spain}
\affiliation{ICREA, Instituci\'o Catalana de Recerca i Estudis Avan\c{c}ats,\\
Passeig de Llu\'{\i}s Companys 23, 
08010 Barcelona, Spain}

\author{Matthew Kirk}
\email{matthew.j.kirk@durham.ac.uk}
\affiliation{Institute for Particle Physics Phenomenology and Department of Physics, Durham University, Durham DH1 3LE, UK}

\begin{abstract}
The deficits w.r.t.~the Standard Model predictions in $B\to K^{(*)}\mu^+\mu^-$ and $B_s\to \phi\mu^+\mu^-$ have recently been substantiated by new analyses from LHCb and CMS, as well as improved theoretical predictions, resulting in strong statistical evidence for lepton flavour universal new physics effects. In this letter, we investigate the possibility of generating such an effect through CKM unitarity violation: If the true value of $V_{cb}$, extracted from $B\to D^{(*)}\ell\nu$, were larger due to an NP contamination, this would lead, via the assumed CKM unitarity within the SM, to a smaller value of $V_{ts}$ and thus to reduced theory predictions for $b\to s\ell^+\ell^-$ branching ratios. A sizable correction to the $W$-$c$-$b$ vertex is only possible close to the custodial-protected limit in which $Z$-$b$-$s$ couplings are absent. However, in a model-independent SMEFT analysis, we show that even if $C_{Hq}^{(1)}=-C_{Hq}^{(3)}$ is realized at a high scale, additional custodial-breaking effects appear due to RGE.
Therefore, only a fine-tuned region in the parameter space of modified gauge boson couplings, displaying an overcompensation of the CKM unitarity-violating effect, gives a good fit to $b\to s\ell^+\ell^-$ data. The only realistic simple SM extensions to achieve this setup are vector-like quarks. However, we show that they cannot explain the $b\to s\ell^+\ell^-$ anomalies without violating bounds for electroweak precision observables as well as $K^0-\bar K^0$ and $B_s-\bar B_s$ mixing.
\end{abstract}

\maketitle

\section{Introduction}\label{sec:intro}

Flavour-changing neutral current (FCNC) processes constitute precision tests of new physics (NP) due to their loop, CKM and Fermi-constant suppression within the Standard Model (SM). In particular, long-standing anomalies, i.e.~tensions between SM predictions and measurements, exist in \bsll transitions. They date back to the first measurement of the angular observable $P_5^\prime$~\cite{Descotes-Genon:2012isb,Descotes-Genon:2013vna} by LHCb in 2013~\cite{LHCb:2013ghj} in 2013 and were later supported by measurements of $B\to K\mu^+\mu^-$~\cite{ATLAS:2018gqc,Belle:2016fev,LHCb:2014cxe,LHCb:2020gog} and $B_s\to \phi\mu^+\mu^-$~\cite{LHCb:2021zwz}. While for some time the ratios $R(K^{(*)})={\rm Br}(B\to K^{(*)}\mu^+\mu^-)/{\rm Br}(B\to K^{(*)}e^+e^-)$ pointed towards the violation of lepton flavour universality (LFU), in 2023, LHCb reanalysed its data and found consistency with the SM predictions~\cite{LHCb:2022zom,LHCb:2022qnv}. 

Nonetheless, the tensions in total branching ratios and angular observables in \bsll transitions were confirmed by new measurements and improved theory predictions in recent years. In particular, the first lattice calculation of $B\to K \ell^+\ell^-$ over the full $q^2$ range~\cite{Parrott:2022rgu} is more than $4\sigma$ above the corresponding measurements. Furthermore, CMS confirmed both the experimental values for $P_5^\prime$~\cite{CMS:2024atz,CMS:2026zas} and Br$(B_s\to \phi\mu^+\mu^-$~)\cite{CMS:2026mhf,CMS:2026zas}. Finally, LHCb reproduced its own measurements of $B\to K^*\ell^+\ell^-$ with more data~\cite{LHCb:2025mqb}. While these new results are not yet all included in global fits, these analyses already find significances beyond the $5\sigma$ level~\cite{Gubernari:2022hxn,Alguero:2023jeh,Hurth:2025vfx,Ciuchini:2022wbq}.\footnote{{Global fits to $b\to s\ell^+\ell^-$ data are sensitive to form factors and the treatment of non-local hadronic matrix elements, both of which remain under active study~\cite{LHCb:2023gpo,LHCb:2026enw,Capdevila:2025drq,Ciuchini:2026hxy,Isidori:2025dkp}. Nevertheless, it was recently pointed out that, with additional data on inclusive decays, and the constraints they impose on the corresponding exclusive modes can enable disentangling NP from a hadronic explanation~\cite{Alvarez-Cartelle:2026nhp}.}}

On the model-building side, this raises the challenge of explaining these LFU hints for NP, which turns out to be more difficult than previously when LFUV was preferred: Couplings to electrons are subject to LEP bounds~\cite{Greljo:2022jac,ALEPH:2010aa} and LFU $b\to s\ell^+\ell^-$ effects with leptoquarks lead to charged lepton flavour violation~\cite{Crivellin:2017dsk}. The attempts for explanations boil down to three main options~\cite{Capdevila:2023yhq}: 1) Off-shell photon penguin with a tau loop~\cite{Crivellin:2018yvo}, induced either via the $U(1)$~\cite{Calibbi:2015kma,Barbieri:2016las,DiLuzio:2017vat,Calibbi:2017qbu,Bordone:2017bld,Blanke:2018sro,Crivellin:2018yvo}, the $S_2$~\cite{Crivellin:2022mff} or the combination of $S_1$ and $S_3$ leptoquarks~\cite{Crivellin:2017zlb,Crivellin:2019dwb,Gherardi:2020qhc} 2) Off-shell photon penguin with charm quarks (diquarks~\cite{Crivellin:2023saq} or charged Higgses~\cite{Iguro:2018qzf,Crivellin:2019dun,Crivellin:2023sig}) 3) LFU $Z^\prime$~\cite{Gauld:2013qba,Buras:2013dea}. While all of them bring theory and experiment closer, a full explanation is difficult, in particular because of the constraints from $B_s-\bar B_s$ mixing~\cite{DiLuzio:2017fdq,DiLuzio:2019jyq}. 

In this work, we want to consider a novel possibility for obtaining an LFU effect in $b\to s\ell^+\ell^-$ transitions: the modification of the $W$-$c$-$b$ interaction leading to CKM unitarity violation. Such a modification is transferred to $V_{ts}$ via the assumed CKM unitarity within the SM (as the direct determination of $V_{ts}$~\cite{ATLAS:2024ojr,CMS:2020vac} is not competitive in precision) and would thus lead to an overestimation of the predictions \bsll rates as well as $B_s-\bar B_s$ mixing, which both involve $V_{ts}$. While the former effect has the potential to explain the anomalies, the latter is welcome to alleviate the constraints on direct NP contributions which arise in any NP model addressing the \bsll anomalies.

We will first examine the possibility of CKM unitarity violation by studying modified $W$-$b$-$c$ couplings in the SM Effective Theory (SMEFT) in Sec.~\ref{Sec:2}. In UV-complete models, such a violation of CKM unitarity arises already at tree-level from additional quarks, such that this has the potential to generate the dominant effect in \bsll transitions. While a chiral $4^{\rm th}$ generation is excluded by the combination of Higgs signal strength measurements and flavour observables~\cite{Eberhardt:2012ck,Eberhardt:2012gv}, vector-like quarks (VLQs) with masses above $\approx 1$\,TeV are compatible with LHC searches~\cite{Benbrik:2024fku,CMS:2024bni} and we will consider them in Sec.~\ref{sec:vlq}.

\section{Effective Field Theory Analysis}
\label{Sec:2}

The effective Lagrangian governing $b \to s \ell^+ \ell^-$ transitions in the SM is commonly defined as
\begin{equation}
\Lagrangian = \frac{4 G_F \V{tb} \V*{ts}}{\sqrt{2}} \frac{\alpha}{4\pi}
\begin{aligned}[t]
    &\left[ C_9^\ell (\bar s \gamma^\mu P_L b)(\bar \ell \gamma_\mu \ell) \right.
    \\
    &\left. + C_{10}^\ell (\bar s \gamma^\mu P_L b)(\bar \ell \gamma_\mu \gamma^5 \ell) \right.
    \\
    &\left. + C_7 \frac{m_b}{e} (\bar s \sigma^{\mu \nu} P_R b) F_{\mu \nu} \right] \,.
\end{aligned}
\end{equation}
Here, we omitted operators with a right-handed $b-s$ current, whose contributions are not favored by the global fit, as well as scalar operators that dominantly affect $B_s\to \mu^+\mu^-$. Furthermore, as the measurements of muon vs electron lepton flavour universality are now consistent with the SM predictions, we assume that $C_{9,10}^e=C_{9,10}^\mu=C_{9,10}$.

\begin{figure}
\includegraphics[width=0.9\linewidth]{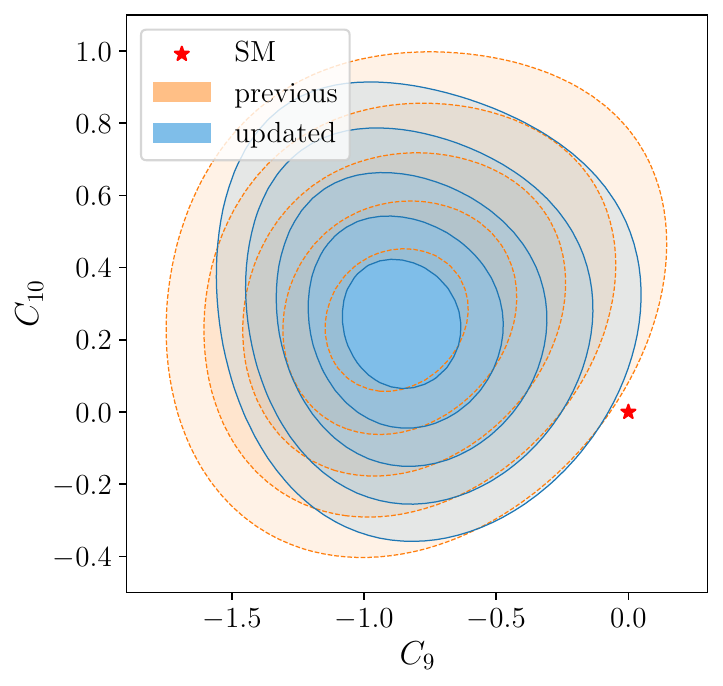}
\caption{Updated fit, including the new LHCb and CMS results, in the $C_9$, $C_{10}$ plane assuming LFU NP (see main text for details).
}
\label{fig:C9-C10-update}
\end{figure}

In light of recent measurements of $b\to s\ell^+\ell^-$ observables by LHCb~\cite{LHCb:2025mqb,LHCb:2025pxz} and CMS~\cite{CMS:2024syx,CMS:2024atz,CMS:2026mhf}, we update the global fit,\footnote{which previously only included the experimental results from Refs.~\cite{Aaij:2014pli,Aaij:2016flj,CMS:2017ivg,Aaij:2020nrf,Aaij:2021pkz,Aaij:2015xza,Lees:2013nxa,CDF:2012qwd,LHCb:2018rym,Aaij:2018gwm,Aaboud:2018krd,LHCb:2020gog,Khachatryan:2015isa,LHCb:2021xxq,Aaij:2020nol,Aaij:2020umj,CMS:2022mgd,LHCb:2021vsc} -- a complete list of the observables and experimental data used in our updated fit can be found in the ancillary file attached to this arXiv submission.} using \flavio~\cite{Straub:2018kue} v2.7.1 and \wilson~\cite{Aebischer:2018bkb} v2.5.2.
The results are shown in Fig.~\ref{fig:C9-C10-update} -- the best fit corresponds to $C_9 = -0.87$, $C_{10} = 0.24$, with the significance of this two-dimensional best fit standing at \qty{5.4}{\sigma}, corresponding to a $\Delta \chi^2$ of $\approx 33$.\footnote{Note that the smaller significance compared to Ref.~\cite{Hurth:2025vfx} is mostly due to the exclusion of the bins closest to the $J/\Psi$ resonance.} 

\begin{figure}
\includegraphics[width=0.85\linewidth]{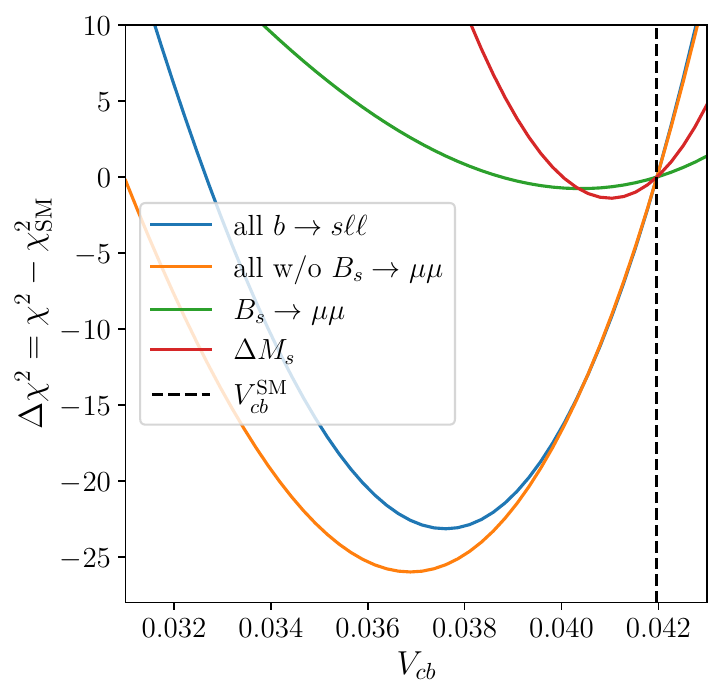}
\caption{ $\Delta\chi^2$ of our $b \to s \ell^- \ell^+$ fit as a function of $\V{cb}$, with the SM value \cite{Finauri:2023kte} marked as a dashed line. A shift in $V_{cb}$ can (only) improve the $\chi^2$ by around 23 due to $B_s\to \mu^+\mu^-$.}
\label{fig:chi2_Vcb_shift}
\end{figure}

Let us now turn to the possibility that this preference for NP could be explained, at least partially, by CKM unitarity violation. For this, it is important to recall that the SM short-distance part of $b \to s \ell^+ \ell^-$ amplitudes is proportional to the product $\V{tb} \V*{ts}$. This combination of CKM elements is usually obtained by assuming CKM unitarity from the measurement of $\V{cb}$ from semi-leptonic $B$ decays, i.e.~$\V{ts}=-\V{cb}$ (up to corrections of the size $\lambda^4$, where $\lambda$ is the Wolfenstein parameter). Hence, if BSM effects alter the extraction of $\V{cb}$, this leads to an indirect shift in FCNC predictions: In case the measured CKM element contains a constructive NP effect, one would overestimate $\V{cb}$ and thus $V_{ts}$ and in turn, the SM prediction. This could (partially) reconcile the theory with the measurements, as can be seen in \cref{fig:chi2_Vcb_shift} where we show the $\Delta\chi^2$ as a function of the CKM element $\V{cb}$: A 10\% shift would provide a reduction of the total $\chi^2$ similar to direct BSM contributions shown in Fig.~\ref{fig:C9-C10-update}. While such a shift of the CKM element $V_{cb}$, and thus of the extracted value of $V_{ts}$ would also impact other observables such as $B_s-\bar B_s$ mixing (see \cref{fig:chi2_Vcb_shift}), in UV-complete models these observables will also receive direct NP effects which in general break this correlation (as we will see for the case of vector-like quarks). 

Let us now study the impact of imposing $SU(2)_L$ gauge invariance, necessarily present in NP models realized above the electroweak breaking scale, by considering the SMEFT Lagrangian
\begin{equation}
\Lagrangian_\text{SMEFT} = 1/\Lambda^2\sum_i C_i Q_i\,,
\end{equation}
for dimension-6 operators in the Warsaw basis~\cite{Grzadkowski:2010es}, where $\Lambda$ is the NP scale for which we will use 2\,TeV in the following. The most important operators are
\begin{equation}
\begin{aligned}
    \OHqthree &= (H^{\dagger} i\overset{\leftrightarrow}{D_{\mu}^I} H)(\bar{q} \tau^I \gamma^{\mu} P_L q) \,, \textrm{ and }
    \\
    \OHqone   &= (H^{\dagger} i\overset{\leftrightarrow}{D_{\mu}} H)(\bar{q} \gamma^{\mu} P_L q) \,,
\end{aligned}
\end{equation}
where $\overset{\leftrightarrow}{D_{\mu}^I}$ and $\overset{\leftrightarrow}{D_{\mu}}$ are defined in \cref{eq:defn_left_right_derivative}, and $q^T_i = \begin{pmatrix}(V^\dagger u_{L})_{i} & d_{L, i} \end{pmatrix}$ is the left-handed quark doublet in the down-aligned basis.
For notational simplicity, in the rest of this work we write the coefficients $C_{Hq}^{(3,1)}$ of these operators as $C^{(1,3)}$, respectively. The explicit definitions of the remaining relevant operators can be found in the appendix.

At tree-level, only the operator $\OHqthree$ leads to modified (left-handed) $W$ couplings
via
\begin{align}
\label{eq:deltaVcb}
V_{ij}^{\rm eff} &= V_{if} \left( \delta_{fj}+ v^2 C^{(3)}_{fj} / \Lambda^2 \right)\,.
\end{align}
Here, $v\simeq 246$\,GeV and $V^{\rm eff}_{if}$ is the effective CKM element as extracted e.g.~from semi-leptonic $d_f\to u_i\ell\nu$ decays, while $V$ (on the right-hand side of the equation) is the unitary CKM matrix of the SM.\footnote{We do not consider right-handed currents induced by $\OHud$ in this work. This operator leads to opposite effects in the determination of $V_{cb}$ from $B\to D\ell \nu$ and $B\to D^*\ell\nu$~\cite{Crivellin:2009sd}, and thus cannot explain the discrepancy with the inclusive determination~\cite{Crivellin:2014zpa}, as the latter is only slightly affected by a right-handed $W$ coupling, while both exclusive determinations are in agreement with each other (see e.g.~Ref.~\cite{HeavyFlavorAveragingGroupHFLAV:2024ctg} for a review on the status of the $V_{cb}$ determinations). Therefore, it has only a limited impact on extracting the value of $V_{cb}$ from data.}

To obtain the desired $O(10\%)$ decrease in $V_{cb}$ w.r.t.~the SM via $C^{(3)}_{22}$, one would need couplings of $O(10)$ for $\Lambda=2$\,TeV. This would not only put one in the non-perturbative regime (even for tree-level NP), but also conflict with LHC bounds~\cite{Eboli:2025vks}. However, $\CHqthreebracket_{23}$ leads to a $1/V_{cb}$ (relatively) enhanced effect, such that for $\Lambda=2$\,TeV, couplings smaller than unity are sufficient. CKM unitarity violation in this form leads to a simple rescaling of the SM prediction for $b\to s\ell^+\ell^-$ decays via
\begin{equation}
C_{9,10}^{\rm SM}\rightarrow C_{9,10}^{\rm SM} \left( 1 - \frac{v^2 }{\V{cb}} \frac{C^{(3)}_{23}}{\Lambda^2} \right) \equiv C_{9,10}^{\rm SM} \left( 1 + \frac{\delta \V{cb}}{\V{cb}} \right) \,.\label{C9CKMshift}
\end{equation}

However, the possible size of $C^{(3)}_{23}$ is strongly constrained experimentally, as it leads to tree-level flavour-violating neutral currents induced via a $Z$-$s$-$b$ coupling. At first glance, a custodial symmetry protecting modifications of the $Z$ boson couplings to down-type quarks, enforced through $\CHqone = -\CHqthree$, would appear to restrict the NP effects to a shift in \V{cb}. However, the SMEFT~\cite{Jenkins:2013zja,Jenkins:2013wua,Alonso:2013hga} and WET RGEs~\cite{Aebischer:2017gaw,Jenkins:2017dyc}, induce $b$-$s$ FCNCs unavoidably: we can write the BSM contribution to the Wilson coefficients at the $B$-meson scale $\mub = \qty{4.8}{\GeV}$ in terms of $C^{(1)}_{23}$ and $C^{(1)}_{23}$ as
\begin{equation}
\begin{aligned}
C_9 (\mub) &\approx \left(33 C^{(3)}_{23}+ 24 C^{(1)}_{23}\right)\frac{\qty{1}{\TeV^2}}{\Lambda^2}\,,
\\
C_{10}  (\mub) &\approx -20\left({30} C^{(3)}_{23} + {27}C^{(1)}_{23} \right)\frac{\qty{1}{\TeV^2}}{\Lambda^2}\,.
\end{aligned}
\label{eq:C9_C10_SMEFT_VLQs}
\end{equation}
Here we have dropped SMEFT coefficients which contribute less than 1\% relative to the leading one. Note that we used $\muNP = \qty{2}{\TeV}$ for the high scale, but the dependence on it is only logarithmic, as we factored out $1/\Lambda^2$. This gives our first no-go theorem: {\bf One cannot explain the \bsll anomalies exclusively via CKM unitarity violation, i.e.~a modified $W$-$c$-$b$ coupling only.}

From Eq.~(\ref{eq:C9_C10_SMEFT_VLQs}), we see clearly that, even if a BSM model produces the pattern $C^{(1)}_{23} = -C^{(3)}_{23}$ at the high scale, i.e.~avoiding tree-level effects in $Z$-$b$-$s$ couplings and in turn generating $|C_{10}|\gg |C_9|$ at the NP scale, model-independent RGE and matching effects induce this unavoidably. Turning to our fit (\cref{fig:C9-C10-update}) we see that the data requires $C_9 < 0$ with $|C_{10}| \ll |C_9|$. As such, at least the two conditions
\begin{enumerate}
    \item $\delta V_{cb} < 0 \Rightarrow C^{(3)}_{23} > 0$ (from \cref{eq:deltaVcb})
    \item $C_9 < 0 \Rightarrow C^{(1)}_{23} \lesssim -1.4 C^{(3)}_{23}$ (from \cref{eq:C9_C10_SMEFT_VLQs})
\end{enumerate}
are necessary (if not necessarily sufficient) for a good explanation of data. Examination of \cref{eq:C9_C10_SMEFT_VLQs} however shows that these two conditions lead to $C_{10} \lesssim -22 C_9$. This gives our second no-go theorem: {\bf One cannot explain the \bsll anomalies through a constructive combination of CKM unitarity violation and direct contributions to $C_9$ without inducing $|C_{10}| \gg |C_9|$.}

\begin{figure}
    \centering
\includegraphics[width=0.95\linewidth]{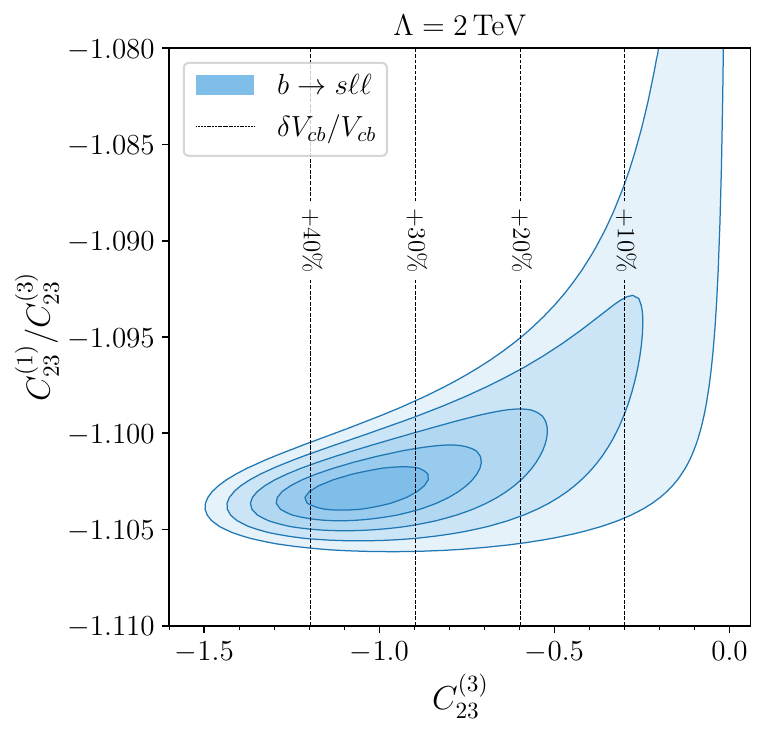}
    \caption{Preferred region from the global \bsll fit in the $C^{(3)}_{23}$--$ C^{(1)}_{23} / C^{(1)}_{23}$ plane along with contour lines denoting the relative shift in $V_{cb}$.} 
    \label{fig:SMEFT}
\end{figure}

This situation is illustrated in Fig.~\ref{fig:SMEFT}. Here, one can see that only a positive shift in \V{cb} (i.e.~with the wrong sign to reduce the theory predictions for $b\to s\ell^+\ell^-$ transition), overcompensated by the (direct) effect of the $Z$-$b$-$s$ coupling gives a good fit to data, manifesting itself as a small fine-tuned region where $C^{(1)}_{23}\approx -1.1C^{(3)}_{23}$ and $\CHqthree_{23} \approx -1$ gives a good fit to data. This region can be most easily understood by defining effective Wilson coefficients
\begin{equation}
\label{eq:C910eff}
    C_{9,10}^\textrm{eff} = C_{9,10}^\textrm{SM} \frac{\delta \V{cb}}{\V{cb}} + C_{9,10} \left( 1 + \frac{\delta \V{cb}}{\V{cb}} \right) \,.
\end{equation}
that include the indirect effect via the shift in $\V{cb}$ from the CKM unitarity violation using Eq.~(\ref{C9CKMshift}). However, a positive shift in $\V{cb}$ of only 10\% would result in an enhancement of the SM prediction for $B_s-\bar B_s$ mixing by 20\%, which is borderline allowed~\cite{Albrecht:2024oyn}. Thus, the question arises if this setup can be realised in a UV-complete model without being in conflict with other measurements because additional direct effects, not just in $B_s-\bar B_s$ mixing but e.g.\ $b\to s\gamma$, arise unavoidably. 

\section{VLQs}
\label{sec:vlq}

First of all, note that the region in the $C^{(3)}_{23}$--$ C^{(1)}_{23} / C^{(1)}_{23}$ plane, which gives a good fit to \bsll data in Fig.~\ref{fig:SMEFT} is close to the custodial limit, avoiding large effects in $Z$-$b$-$s$ couplings. Therefore, we need a UV completion that results in values for the Wilson coefficients close to $C^{(1)}=-C^{(3)}$. The small deviations from this limit, needed to give a good fit to data, can be due to higher-order matching corrections or adding additional particles generating subleading effects. 

The only single-particle extensions of the SM that give rise to the Wilson coefficient $C^{(3)}_{ij}$ at tree level are the four VLQ representations $U$, $D$, $T_1$, $T_2$ (see the supplementary material~\ref{sec:vlq_reps}), or an $SU(2)_L$ triplet, hypercharge zero vector boson~\cite{deBlas:2017xtg}. However, the latter is not renormalizable on its own and building a model with dominantly flavour-violating effects is challenging. Therefore, we study the VLQ option, which gives rise to
\begin{align}
\label{eq:VLQ_Hq_SMEFT_matching}
C^{(3)}_{ij} &= 
    -\frac{\xi_{i}^{U} \xi_{j}^{U}}{4 M_U^2 }
    - \frac{\xi_{i}^{D} \xi_{j}^{D}}{4 M_D^2}
    + \frac{ \xi_{i}^{T_1} \xi_{j}^{T_1} }{ 16M_{T_1}^2 }
    + \frac{ \xi_{i}^{T_2} \xi_{j}^{T_2} }{ 16M_{T_2}^2 }
\,, \\
C^{(1)}_{ij} &= 
    \phantom{+} \frac{\xi_{i}^{U} \xi_{j}^{U}}{4 M_U^2 }
    - \frac{\xi_{i}^{D} \xi_{j}^{D}}{4 M_D^2}
    - \frac{3 \xi_{i}^{T_1} \xi_{j}^{T_1} }{ 16M_{T_1}^2 }
    + \frac{3 \xi_{i}^{T_2} \xi_{j}^{T_2} }{ 16M_{T_2}^2 }
\end{align}
at tree-level. From this we can estimate that for the desired $O(10\%)$ decrease in $V_{cb}$ we need $\xi^{(X)}_{2,3} \sim 1-2$ for a VLQ mass of $2\,$TeV.

\begin{figure}
    \includegraphics[width=0.9\linewidth]{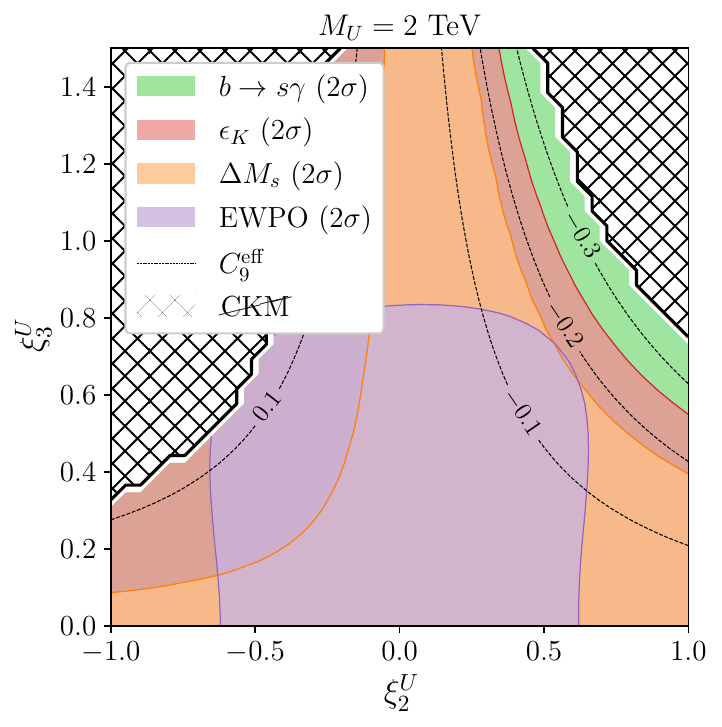}
    \caption{Preferred regions for the VLQ $U$ in the $\xi_2^U$--$\xi_3^U$ plane for $M_U=2$\,TeV. The hatched region ($\cancel{\textrm{CKM}}$) shows where the CKM extraction (as detailed in the supplementary material \ref{ckm_treatment}) fails. Note that $B_s-\bar B_s$ mixing and EWPO do not allow for a full explanation of $b\to s\ell^+\ell^-$ data.}
    \label{fig:U_VLQ}
\end{figure}

First, for the case of a single VLQ, only $U$ results in custodially protected $Z$ couplings to down-type quarks at tree-level. The preferred and excluded regions, together with the effect in $C_9^{\rm eff}$ (defined in \cref{eq:C910eff}), are shown in Fig.~\ref{fig:U_VLQ}. Here, we set the masses of the VLQs to \SI{2}{\TeV}, consistent with the bounds for third-generation VLQs from ATLAS and CMS~\cite{ATLAS:2018ziw,ATLAS:2021ibc,ATLAS:2021ddx,CMS:2024bni}. For $U$, the one-loop matching corrections can shift $C^{(1)}_{23}/C^{(3)}_{23}$ to no more than $-1.03$.\footnote{The fit includes the one-loop matching of the VLQ model on the SMEFT using \mmeft~\cite{Carmona:2021xtq} (see the supplementary material~\ref{app:loop-matching}) and the one-loop SMEFT-WET matching~\cite{Dekens:2019ept}, and has been conducted using \smelli~\cite{Aebischer:2018iyb,Stangl:2020lbh} v2.4.3. Minimal modifications were made to add the new data from LHCb and CMS, alongside altering the theory prediction for $\Delta M_s$ to use the combination of non-perturbative inputs from Refs.~\cite{FermilabLattice:2016ipl,Dowdall:2019bea,King:2019lal} performed in Ref.~\cite{Greljo:2022jac}.}
Beyond that, adding another VLQ to achieve $C^{(1)}_{23}/C^{(3)}_{23}\approx-1.1$ would not result in significantly larger allowed values of $C_9$. The reason is that only relatively small couplings of the additional VLQ are needed to perturb $C^{(1)}_{23}/C^{(3)}_{23}\approx-1.03$ to $\approx-1.1$. Therefore, the effect of adding the additional VLQ will leave the $B_s-\bar B_s$ mixing bounds nearly unchanged, thus preventing larger effects in $C_9$.

\begin{figure*}
    \includegraphics[width=0.9\linewidth]{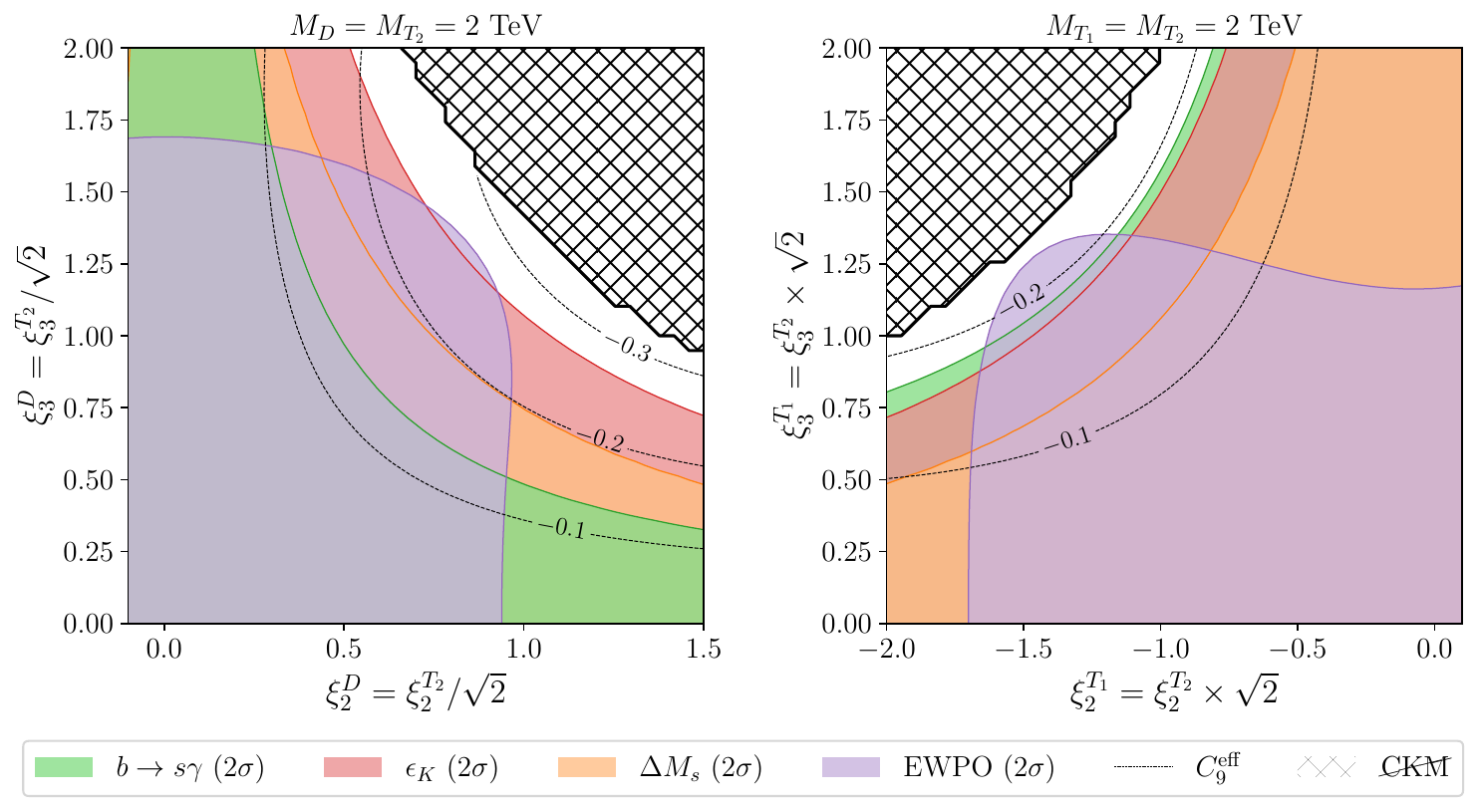}
    \caption{Preferred regions for the model with the VLQs $D$ and $T_2$ in the $\xi_2^D$--$\xi_3^D$ plane (left) and $T_1$ and $T_2$ in the $\xi_2^{T_1}$--$\xi_3^{T_1}$ plane (right) for \qty{2}{\TeV}. In both cases, the relative size of the VLQ couplings is fixed to ensure custodial protection. The hatched region ($\cancel{\textrm{CKM}}$) shows where the CKM extraction (as detailed in the supplementary material \ref{ckm_treatment}), fails. Note that $B_s-\bar B_s$ mixing and EWPO do not allow for a full explanation of $b\to s\ell^+\ell^-$ data.}
    \label{fig:X+T2_VLQs}
\end{figure*}

For the case of adding two VLQs to the SM, there are only two options to obtain custodially protected $Z$ down-quark interactions (disregarding the possibility of a maximal complex phase between the couplings of the different VLQs): $D$ plus $T_2$ or $T_1$ plus $T_2$, for which $C^{(1)}=-C^{(3)}$ is obtained for $\xi^{T_2}_i=\sqrt{2} \xi^{D}_i$ and $\xi^{T_1}_i=\sqrt{2} \xi^{T_2}_i$, respectively. From Fig.~\ref{fig:X+T2_VLQs}, one can see that the situation with two VLQs leads to the same problems as in the case of the single VLQ $U$: $C_9$ cannot be negative enough to explain $b\to s\ell^+\ell^-$ data. While we refrained from showing the size of $C_{10}$ in this plot, which turns out to be too large, this could be mitigated by adding another VLQ as a small perturbation. But again, this could not lead to a full explanation of $b\to s\ell^+\ell^-$ data, as $C_9^{\rm eff}$ would still be small, as the constraints from  $B_s-\bar B_s$ mixing remain nearly unchanged. Furthermore, even if one canceled the effect in  $B_s-\bar B_s$ mixing by some other kind of NP, $\epsilon_K$ would still exclude large effects.

Therefore, we find our third no-go theorem:
{\bf There is no simple UV-complete model which can lead to the overcompensation solution with $C^{(3)}_{23}\approx -0.9 C^{(1)}_{23}\approx -1$ to explain $b \to s \ell^+ \ell^-$ data via modified EW gauge boson couplings to quarks.}

\section{Conclusions}
\label{sec:conclu}

In this article, we examined CKM unitarity violation in the form of potential shifts to the extraction of \V{cb} from NP, and the subsequent effect on the theory predictions for \bsll. Within the SM, the unitarity of the CKM matrix is assumed, and the CKM element \V{ts}, which enters \bsll theory predictions, is typically determined from \V{cb}, extracted from $B \to D \ell \nu$ decays. As such, an overestimation of \V{cb}, because of unaccounted BSM effects, would lead to a corresponding overestimation of \bsll observables, as seen in current data.

Starting with an SMEFT analysis, we showed that all high-scale NP that modifies the left-handed $W$-$c$-$b$ coupling inevitably leads to a modification of the $Z$-$b$-$s$ interaction, and hence a BSM contribution to $C_{9,10}$, at the $B$ meson scale. The model-independent structure of the renormalisation group equations generically gives rise to the pattern $|C_{10}| \gg |C_9|$, except through careful fine-tuning of the SMEFT coefficients close to the custodial limit ($C^{(1)}_{23}=-C^{(3)}_{23}$), where the CKM effect is overcompensated by the effect of the $Z$-$b$-$s$ coupling.

We then identified VLQs as the most viable UV completion, as they naturally give tree-level modifications of EW gauge boson couplings, and studied three models which can naturally lead to the high-scale custodial protection of diagonal $Z$ couplings. In this custodial limit, we found that current bounds from $B_s-\bar B_s$ mixing and electroweak precision observables (EWPO) limit the effect in $C_9$ to be less than one-quarter of the necessary size to fully explain the anomaly. In light of this, any small perturbation, either through additional VLQs or higher-order matching corrections, cannot enhance the 
\bsll effect and at the same time escaping the tight experimental bounds from other observables.

Taken together, in our analysis we demonstrated three no-go theorems related to an explanation of the \bsll anomalies through CKM unitarity violation:
\begin{enumerate}
    \item While in the SMEFT $O^{(3)}_{23}$ leads to CKM unitarity violation, this operator alone cannot result in sizable effects as in results in modified $Z$-$s$-$b$ couplings.
    \item Only a fine-tuned region around $C^{(3)}_{23}\approx-C^{(1)}_{23}\approx -1$, where the effect in $V_{cb}$ has the wrong sign but is overcompensated by the effect of the $Z$-$s$-$b$ coupling can explain $b\to s\ell^+\ell^-$ data. 
    \item The only realistic SM extension to achieve this setup are VLQ. However, it is not possible to obtain large enough values of $C^{(1,3)}_{23}$, and in turn $C_9$, without violating other experimental bounds, in particular $B_s-\bar B_s$ mixing and EWPO.
\end{enumerate}

\begin{acknowledgments}
A.\ C.\ thanks Zurab Berezhiani for useful discussions.
M.\ K.\ thanks Peter Stangl and Aleks Smolkovic for helpful advice regarding implementation of experimental measurements in \flavio.
\end{acknowledgments}

\bibliographystyle{JHEP}
\bibliography{BIB}

\onecolumngrid

\setcounter{equation}{0}
\setcounter{figure}{0}
\setcounter{table}{0}
\setcounter{section}{0}
\makeatletter
\renewcommand{\theequation}{S.\arabic{section}.\arabic{equation}}
\renewcommand{\thefigure}{S.\arabic{figure}}
\renewcommand{\thetable}{S.\arabic{table}}
\renewcommand{\thesection}{\arabic{section}}

\begin{center}
 \large{\bf Supplemental Material}
 \end{center}

\section{SMEFT operators}

Here we give the definitions of the SMEFT operators relevant for flavour and electroweak precision observables according to Ref.~\cite{Grzadkowski:2010es}. $p$, $r$, $s$ and $t$ are flavour indices, while color as well as $SU(2)_L$ indices are contracted within the bi-linears and $T^A$ stands for the generators of $SU(3)_C$.

\paragraph*{Modified gauge boson couplings:}
The operators generating modified gauge boson couplings to quarks after EW symmetry breaking are
\begin{align}
Q_{H q}^{(1)ij} = (H^{\dagger}i\overleftrightarrow{D_{\mu}}H)(\bar{q}_i\gamma^{\mu}q_j)\,,
\qquad
Q_{H q}^{(3)ij} = (H^{\dagger}i\overleftrightarrow{D_{\mu}^I}H)(\bar{q}_i\tau^I\gamma^{\mu}q_j)\,,
\label{eq:HiggsqOperators}
\end{align}
with
\begin{equation}
\label{eq:defn_left_right_derivative}
\overleftrightarrow{D_{\mu}} = (D_{\mu} - \overset{\leftarrow}{D}_{\mu})\,,\qquad \overleftrightarrow{D_{\mu}^I} = (\tau^I D_{\mu} - \overset{\leftarrow}{D}_{\mu}\tau^I)\,.
\end{equation}
and the covariant derivative 
\begin{equation}
   D_{\mu} = \partial_{\mu}
   + i g_1 Y B_{\mu}
   + i g_2 S^I W^I_{\mu}
   + i g_s T^A G^{A}_{\mu}\,.
   \label{covariantD}
\end{equation}
Here $T^A = \frac{1}{2}\lambda^A$ and $S^I = \tau^I$, with $\lambda^A$ and $\tau^I$ the Gell-Mann and the Pauli matrices respectively.

It is useful to write explicitly the modifications of the \PW and \PZ couplings to left-handed quarks (after EW symmetry breaking) as a function of the SMEFT coefficients:
\begin{align}
\delta \Lagrangian_{W,Z} =
&- v^2 \frac{g}{\sqrt{2}} W^+_\mu \, \bar{u}_i \gamma^\mu \left( \left[ V \cdot \CHqthree \right]_{ij} P_L
\right) d_j + \hc \label{eq:W_modifications}
\\ 
&- v^2 \frac{g}{2 c_W} Z_\mu \, \bar{u}_i \gamma^\mu \left( \left[ V \cdot \left\{ \CHqthree - \CHqone \right\} \cdot V^\dagger \right]_{ij} P_L
\right) u_j  \label{eq:Zuu_modifications}
\\
&- v^2 \frac{g}{2 c_W} Z_\mu \, \bar{d}_i \gamma^\mu \left( \left[ \CHqthree + \CHqone \right]_{ij} P_L
\right) d_j \,,  \label{eq:Zdd_modifications}
\end{align}
where $V$ is the CKM matrix and $v\simeq 246$\,GeV.

\section{Vector-like quark representations}
\label{sec:vlq_reps}

Defining VLQs as heavy fermions which are triplets of $SU(3)_C$ and that can mix with the SM quarks after EW symmetry breaking, i.e.\ fermions which can have couplings to the SM Higgs and a SM quark, there are seven possible representations (under the SM gauge group $SU(3)_C\times SU(2)_L\times U(1)_Y$) of VLQs, given in \cref{tab:VLQrepresentations}, of which only four ($U$, $D$, $T_1$, $T_2$) are relevant for this work.
The kinetic and mass terms\footnote{Note that mass terms such as $m_{i}^U \bar U {u_i}$ can always be removed by a field redefinition, such that the kinetic terms and the mass terms take the diagonal form shown in Eq.~\eqref{eq:VLQ_lagrangian}.}
 are
\begin{equation}
\label{eq:VLQ_lagrangian}
\mathcal{L} = \sum_{F} \bar{F} \left( i  \slashed{D}  - M_F \right) F\,,
\end{equation}
where $F=\{U,D,T_1, T_2\}$.
Here $(S^I)_{jk}$ are $0$ and $-i \epsilon_{Ijk}$ for the $SU(2)_L$ singlet and triplet representations respectively.
The (generalized) Yukawa couplings are encoded in the Lagrangian
\begin{align}
{{\cal L}_{}} = {{\cal L}^H_{qq}}
+ {{\cal L}^H_{q \text{VLQ}}}
\,,
\end{align}
where the first term contains the SM Yukawa couplings
\begin{align}
\begin{aligned}
 -{{\cal L}^H_{qq}} =&\, Y^u_{ij} \bar{q}_i \tilde{H} u_j + Y^d_{ij} \bar{q}_i H d_j + \text{h.c.}\,,
 \end{aligned}
\end{align}
and the second term the Higgs interactions with vector-like and SM quarks
\begin{align}
- {{\cal L}^H_{q \text{VLQ}}} &=
\xi _{i}^U{{\bar U}}{{\tilde H}^\dag }{q_i} + \xi _{i}^D{{\bar D} }{H^\dag }{q_i}
+\frac{1}{2}\xi _{i}^{{T_1}}{H^\dag }\tau \cdot{{\bar T}_{1}}{q_i} + \frac{1}{2}\xi _{i}^{{T_2}}{{\tilde H}^\dag }\tau \cdot{{\bar T}_{2}}{q_i} + \hc \,,
\end{align}
Here $i,j = \{1,2,3\}$ are flavour indices and $\tau\cdot\bar T=\sum_I \tau^I \bar T^I$.

\begin{table}[t]
\centering
\renewcommand{\arraystretch}{1.2}
\begin{tabular}[t]{l |  c c c c | c c c c c c c }
\toprule
           & $u$ & $d$ & $q$ & $H$ & $U$ & $D$ & $Q_1$ & $Q_5$ & $Q_7$ & $T_1$ & $T_2$ \\
\midrule
$SU(3)_C$ &   3  &  3   &   3   &   1   &   3   &   3   &   3   &   3   &   3   &   3   &   3   \\
$SU(2)_L$ &   1  &  1   &   2   &   2   &   1   &   1   &   2   &   2   &   2   &   3   &   3   \\
$U(1)_Y$  &\sfrac{2}{3}& \sfrac{-1}{3} & \sfrac{1}{6} & \sfrac{1}{2} & \sfrac{2}{3} & \sfrac{-1}{3} & \sfrac{1}{6} & \sfrac{-5}{6} & \sfrac{7}{6} & \sfrac{-1}{3} & \sfrac{2}{3} \\
\bottomrule
\end{tabular}
\caption{Representations of the SM Higgs, the SM quarks and the VLQs under the SM gauge group. 
}
\label{tab:VLQrepresentations}
\end{table}

\section{CKM treatment}
\label{ckm_treatment}

Since we modify $W$-quark couplings at tree-level, the determination of the ``correct'' CKM matrix elements is non-trivial. In \smelli, the method of Ref.~\cite{Descotes-Genon:2018foz} is implemented, which fixes the four inputs needed to determine the three CKM angles and its complex phase using four observables, fully taking into account both the SM and new physics contributions, at each point in parameter space.
These four observables are $\Gamma(K^+ \to \mu^+ \nu) / \Gamma(\pi^+ \to \mu^+ \nu)$, $\mathrm{Br}(B \to X_c e^+ \nu)$, $\mathrm{Br}(B^+ \to \tau^+ \nu)$, and $\Delta M_d / \Delta M_s$. Once the CKM matrix is fixed, it is then used as input to calculate all the other theory predictions, including the SM predictions. Note that this means there is an input scheme dependence which determines which observables display discrepancies from the SM predictions, because the four used to fix the CKM matrix agree by construction with experiment. Therefore, only the (change in the) global $\Delta \chi^2$ is physical. For sufficiently large NP effects, no consistent extraction can be made, and this is denoted by hatched regions in the figures in \cref{sec:vlq}.

\section{VLQ to SMEFT matching coefficients}
\label{app:loop-matching}

At tree-level, in addition to those given in the main text (\cref{eq:VLQ_Hq_SMEFT_matching}), we also find for the Wilson coefficients of the three Higgs operators
\begin{equation}
\left[C_{uH}\right]_{ij} = \left( \frac{ \xi^U_{i} \xi^{U}_k}{2 M_{U}^2} + \frac{ \xi^{T_1}_{i} \xi^{T_1}_{k}}{4 M_{T_1}^2} + \frac{ \xi^{T_2}_{i} \xi^{T_2}_{k}}{8 M_{T_2}^2} \right) Y^u_{kj}
\,, \qquad
\left[C_{dH}\right]_{ij} = \left( \frac{ \xi^D_{i} \xi^{D}_k}{2 M_{D}^2} + \frac{ \xi^{T_1}_{i} \xi^{T_1}_{k}}{8 M_{T_1}^2} + \frac{ \xi^{T_2}_{i} \xi^{T_2}_{k}}{4 M_{T_2}^2} \right) Y^d_{kj}
\end{equation}

For modified gauge boson couplings, beyond the tree-level matching given in \cref{eq:VLQ_Hq_SMEFT_matching}, there are 1-loop matching contributions which already break the custodial symmetry we impose at tree-level. We only present here the 1-loop matching terms proportional to either the new physics coupling $\xi$, or the SM up-type Yukawa $Y^u$, thus neglecting the down-type and lepton Yukawas as well as the gauge couplings $g_1, g_2$.
For the $U$ VLQ, we find: 
\begin{align}
\left[ C_{Hq}^{(1)} \right]_{ij} &= 
       \frac{\xi_j^U \xi_k^{U} Y_{il}^u Y_{lk}^{u \dagger}}{32 \pi ^2 M_U^2} \left(1 + \ln \frac{\mu^2}{M_U^2} \right)
     + \frac{\xi_k^U  \xi_i^{* U} Y_{kl}^u Y_{lj}^{u \dagger}}{32 \pi ^2 M_U^2} \left(1 + \ln \frac{\mu ^2}{M_U^2} \right)
     - \frac{\xi _j^U \xi _k^U \xi _i^{* U} \xi _k^{* U}}{256 \pi ^2 M_U^2} \left( 17 + 14 \log \frac{\mu ^2}{M_U^2} \right)
\,, \\
\left[ C_{Hq}^{(3)} \right]_{ij} &=
\frac{\xi_j^U \xi_k^U \xi_i^{U} \xi_k^{U}}{256 \pi^2 M_U^2} \left( 9 + 14\ln \frac{\mu ^2}{M_U^2} \right)
\,.
\end{align}
For the $D$ and $T_2$ custodial combination (i.e.\ we already in the below set $\xi^{T_2}_i=\sqrt{2} \xi^{D}_i$) we find:
\begin{align}
\left[ C_{Hq}^{(1)} \right]_{ij} &= 
    \frac{\xi_j^D \xi_k^{D} Y_{il}^u Y_{lk}^{u \dagger}}{256 \pi ^2 M_D^2} \left(3 + 2\ln \frac{\mu^2}{M_D^2} \right)
  + \frac{\xi_k^D  \xi_i^{D} Y_{kl}^u Y_{lj}^{u \dagger}}{256 \pi ^2 M_D^2} \left(3 + 2 \ln \frac{\mu ^2}{M_D^2} \right)
  - \frac{\xi_j^D \xi_k^D \xi_i^{D} \xi_k^{D}}{1024 \pi ^2 M_D^2} \left( 37 + 70 \log \frac{\mu ^2}{M_D^2} \right)
\,, \\
\left[ C_{Hq}^{(3)} \right]_{ij} &=
  - \frac{\xi_j^D \xi_k^{D} Y_{il}^u Y_{lk}^{u \dagger}}{256 \pi ^2 M_D^2} \left(7 + 6\ln \frac{\mu^2}{M_D^2} \right)
  - \frac{\xi_k^D  \xi_i^{D} Y_{kl}^u Y_{lj}^{u \dagger}}{256 \pi ^2 M_D^2} \left(7 + 6 \ln \frac{\mu ^2}{M_D^2} \right)
  + \frac{\xi_j^D \xi_k^D \xi_i^{D} \xi_k^{D}}{1024 \pi ^2 M_D^2} \left( 37 + 70 \log \frac{\mu ^2}{M_D^2} \right)
\,.
\end{align}
For the $T_1$ and $T_2$ custodial combination (i.e.\ we already in the below set $\xi^{T_1}_i=\sqrt{2} \xi^{T_2}_i$), we find:
\begin{align}
\left[ C_{Hq}^{(1)} \right]_{ij} &= 
  - \frac{3\xi_j^{T_1} \xi_k^{{T_1}} Y_{il}^u Y_{lk}^{u \dagger}}{1024 \pi ^2 M_{T_1}^2} \left(3 + 2\ln \frac{\mu^2}{M_{T_1}^2} \right)
  - \frac{3\xi_k^{T_1}  \xi_i^{{T_1}} Y_{kl}^u Y_{lj}^{u \dagger}}{1024 \pi ^2 M_{T_1}^2} \left(3 + 2 \ln \frac{\mu ^2}{M_{T_1}^2} \right)
  + \frac{63\xi_j^{T_1} \xi_k^{T_1} \xi_i^{{T_1}} \xi_k^{{T_1}}}{16384 \pi ^2 M_{T_1}^2} \left( 5 + 6 \log \frac{\mu ^2}{M_{T_1}^2} \right)
\,, \\
\left[ C_{Hq}^{(3)} \right]_{ij} &=
  + \frac{\xi_j^{T_1} \xi_k^{{T_1}} Y_{il}^u Y_{lk}^{u \dagger}}{1024 \pi ^2 M_{T_1}^2} \left(5 + 2\ln \frac{\mu^2}{M_{T_1}^2} \right)
  + \frac{\xi_k^{T_1}  \xi_i^{{T_1}} Y_{kl}^u Y_{lj}^{u \dagger}}{1024 \pi ^2 M_{T_1}^2} \left(5 + 2 \ln \frac{\mu ^2}{M_{T_1}^2} \right)
  - \frac{5\xi_j^{T_1} \xi_k^{T_1} \xi_i^{{T_1}} \xi_k^{{T_1}}}{16384 \pi ^2 M_{T_1}^2} \left( 71 + 82 \log \frac{\mu ^2}{M_{T_1}^2} \right)
\,.
\end{align}

\end{document}